\documentclass[prl,two column,showpacs,preprintnumbers,amsmath,amssymb,revtex4-1]{revtex4-1}
\usepackage{amsfonts, amsmath, bm, bbm, graphicx, epsfig, epstopdf}
\usepackage{dcolumn}
\usepackage{textcomp}
 \usepackage[caption=false]{subfig}
\usepackage{mathtools}
\usepackage{multirow}   

\usepackage{xcolor}
    \usepackage{braket}
\renewcommand\bra[1]{{\langle{#1}|}}
\makeatletter
\renewcommand\ket[1]{%
  \@ifnextchar\bra{\k@t{#1}\!}{\k@t{#1}}%
}
\newcommand\k@t[1]{{|{#1}\rangle}}
\makeatother

\begin{document}
\title{Quantum computation with longlived Rydberg-Landau atoms \\featuring suppressed ionization by the Magnetic Cage}

\date{\today}

\author{Amirhossein Momtaheni, Mohammadsadegh Khazali}
\affiliation{ Department of Physics,  University of Tehran, Tehran, Iran }
\begin{abstract}
Atomic processing units require robust entanglement between individual qubits, typically achieved via excitation to highly interacting Rydberg states. However, short Rydberg lifetimes and ionization risks limit the quantum volume score of the atomic processing units.  Inspired by Landau resonances in alkaline atoms, we introduce Rydberg-Landau (rLandau) states created under a strong magnetic field (2.5 Tesla). These states exhibit significantly extended lifetimes and a magnetic confinement mechanism that prevents ionization, even under intense laser fields. We analyze their wavefunctions, excitation dynamics, dipole transition rules, lifetimes, and interactions, identifying states optimal for high-fidelity quantum operations. This approach simplifies the coherent excitation of long-lived, strongly interacting rLandau circular states akin to Coulombic counterparts, enabling deeper and more complex quantum algorithms.
\end{abstract}

\maketitle

\section{introduction}

Atomic processing units have emerged as a pioneering platform for scalable quantum technology. Implementing quantum algorithms in these units demands precise entanglement of individual atomic qubits, typically achieved by exciting atoms from their non-interacting ground states to highly interacting Rydberg states \cite{Saf10,Blu24,Kha25,KhaTerminal24,Kha23,KhaFermi23,Kha22,Kha21,Kha20,Kha19,Kha15,Wei10}. However, the inherently short lifetimes of these Rydberg states \cite{Bet09} significantly limit the complexity and depth of quantum algorithms that could be run on atomic processors, hindering the full realization of quantum advantage. The error correction could not be a solution in the current era of noisy intermediate-scale quantum (NISQ) devices due to the significant overhead in physical qubit numbers \cite{Gid21} and highly complex gate operations between the delocalized logical degrees of freedom \cite{Fow12}. Addressing this challenge necessitates the engineering of new atomic orbitals that significantly enhance the coherence over the entangling process in atomic units.

One promising approach utilizes circular Rydberg states, whose donut-shaped electron clouds minimize overlap with ground-state wavefunctions, achieving exceptionally long lifetimes on the order of minutes \cite{Ngu18}. Nevertheless, coherent excitation of circular states remains technically demanding  \cite{Sig17,Car20}, often requiring complex, multi-photon excitation schemes involving numerous laser frequencies.

An alternative strategy involves exciting higher principal numbers $n$ to increase the interaction-to-loss ratio. Furthermore, increasing the excitation laser intensity enables faster gate operations in resonant schemes and improves the interaction-to-loss ratio in off-resonant Rydberg dressing protocols \cite{Kha24,Kha18,Kha16,Khaz21,Shi24}. 
Unfortunately, intense laser fields and weak binding of highly excited states significantly raise ionization risks, limiting the practical advantages of this method.

Historically, spectroscopic studies of highly excited alkaline atoms under strong magnetic fields have demonstrated Landau-type resonances extending even into the continuum energy regime \cite{Gar69,Fon78,Fon80,Eco78,Gay80,Cas80}.  Inspired by these phenomena, we propose employing strong magnetic fields to engineer long-lived, highly excited Rydberg-Landau (rLandau) states. In these states, the applied magnetic field of approximately 2.5 Tesla generates a robust quadratic potential in the plane perpendicular to the field direction, generating a Landau-type wavefunction, and effectively creating a magnetic cage that prevents atomic ionization even under strong laser excitation. Concurrently, along the magnetic field axis, the electron's wavefunction retains the characteristic Coulombic features typical of Rydberg states. Finally, we introduce the ultra-long-lived rLandau states that are akin to the circular states in the electron cloud orbital, with the advantage of a simple, coherent excitation scheme compared to the coulombic circular states.

In this work, we study the wavefunction of these novel rLandau states and thoroughly examine their laser excitation dynamics, identifying and classifying dipole transition selection rules. We quantify state lifetimes, and trace back the longevity of the rLandau states to small overlap with the low-lying coulombic states, reduced dipole allowed accessible state, or being at the minimum energy state while maintaining strongly interacting dipoles.

 pinpointing the circular-type doughnut shape wave functions without dipole-allowed decay channels, thereby highlighting their exceptional longevity. Additionally, both dipolar and van der Waals interactions among rLandau states are evaluated. Finally, the paper discusses how the quadratic magnetic field extends the rLandau states over the ionization continuum region in the free field regime, which suppresses the density of states in the Fermi golden rule and mitigates the ionization rate.


Our approach addresses two critical limitations simultaneously: it enables coherent excitation of ultra-long-lived, highly interacting states using significantly fewer lasers compared to Coulombic circular states. It establishes a magnetic confinement mechanism to prevent ionization over the fast gate operation robustly. These advancements pave the way toward high-fidelity quantum operations, enabling the practical implementation of complex, deeply structured quantum algorithms.

\section{Results}
\subsection{Rydberg-Landau Wavefunction}

The Hamiltonian describing hydrogen-like atoms in the presence of a uniform magnetic field exhibits cylindrical symmetry and can be expressed as:
\begin{eqnarray}\label{a14} \nonumber
H &=& -\frac{\hbar^2\nabla^2}{2m}- \frac{e}{2m}(\mathbf{L}+2\mathbf{S})\cdot\mathbf{B}+\xi(r)\mathbf{L}\cdot\mathbf{S}\\
&&+\frac{e^2}{8m}B^2\rho^2-\frac{e^2}{4\pi\epsilon_0 \sqrt{\rho^2+z^2}},
\end{eqnarray}
where \( m \) is the electron mass, $e$ is the magnitude of the electron charge and \( \xi(r) \) characterizes the spin-orbit interaction.

When analyzing the electron’s wavefunction within hydrogen-like Rydberg states, a useful starting point is to approximate its spatial extension using Coulomb eigenstates. In these states, the characteristic orbital radius scales as \(n^2 a_0\), leading to a radial wavefunction size that grows quickly with the principal quantum number \(n\). In the presence of a magnetic field \(B\), the relevant magnetic energy scales—particularly the diamagnetic (quadratic) term—increase in significance at large \(n\) compared to SOC with $\xi(r)\propto r^{-3}$ and coulumb term $\propto r^{-1}$. A practical criterion for identifying when these magnetic terms dominate over spin-dependent effects is \(n^4 B \gg 1\). Under this condition, the magnetic interaction energies become sufficiently large to decouple the electron’s spin \(\mathbf{S}\) from its orbital motion \(\mathbf{L}\), allowing one to neglect both \(\mathbf{S}\) and the spin–orbit coupling \(\mathbf{L}\cdot\mathbf{S}\) to a good approximation.  For example, at n=65 with $B=2.5$T, the diamagnetic term $E_{dia}\approx e^2B^2n^4a_0^2/8m$ is 30 times larger than spin-orbit coupling $E_{SOC}\approx \alpha^2 Ry/n^3$.

Considering the spatial spread of highly excited atoms, the total Hamiltonian in Eq.~\eqref{a14} simplifies to the spin-independent form
\begin{equation}
H = -\frac{\hbar^2 \nabla^2}{2m} -\frac{ e}{2m}\,\vec{L}\cdot \vec{B} + \frac{e^2}{8m}B^2\rho^2 - \frac{\frac{e^2}{4\pi\epsilon_0}}{ \sqrt{\rho^2+z^2}},
\label{Eq_H}
\end{equation}
in which the orbital–magnetic coupling (linear and quadratic Zeeman terms) and the Coulomb potential are the principal contributors to the electronic structure. This assumption greatly reduces theoretical complexity when describing strongly magnetized, highly excited states.

Considering the remaining terms of the potential, the ratio of diamagnetic (quadratic) to linear magnetic terms scales as \( \frac{ea_0^2}{4\hbar m_l} B n^4 \). Under a magnetic field of 2.5 T, for a low-lying angular momentum state $m_l=1$, the linear magnetic term dominates for \( n < 15 \), while the quadratic magnetic term becomes dominant for \( n > 45 \). 
Comparing magnetic and Coulomb potentials, for lower principal quantum numbers (\( n<15 \)), the ratio of linear magnetic to Coulomb terms  $   \frac{e \hbar B / 2m}{e^2 / 4\pi\epsilon_0 r} \propto B n^2$ is negligible. Conversely, for higher excited states (e.g., \( n>65 \)), the quadratic magnetic term dominates the Coulomb term  $\frac{e^2 B^2 r^2 / 8m}{e^2 / 4\pi\epsilon_0 r} \propto B^2 n^6$.
Hence, at typical ground-state levels of rubidium (\( n=5 \)), Coulomb attraction is dominant, whereas, in highly excited states, magnetic confinement prevails.

Spectroscopic studies have shown that under strong magnetic fields, high-\( n \) Rydberg states would be replaced by Landau-type states \cite{Gar69,Fon78,Fon80,Eco78,Gay80,Cas80}. 
In a strong magnetic field \( \mathbf{B} \) oriented along the \( z \) axis, the electron’s motion in the perpendicular plane is significantly modified by the linear and quadratic magnetic terms. These terms provide tight radial (cyclotron-like) confinement and produce energies that can exceed the typical Rydberg energy separation for high Rydberg states. Consequently, the wavefunction in the \((\rho, \phi)\) plane becomes effectively “frozen” into Landau-like modes, with minimal coupling to axial (\(z\)) motion. The electron thus behaves as though it were subject to a two-dimensional harmonic oscillator in the perpendicular plane and a modified one-dimensional Coulomb potential along the \(z\) axis. Under these conditions, it is natural to seek approximate solutions by separating the wavefunction
\[
\Psi(\rho,\phi,z) \;=\; f_{N_z,P}(z)\,Q_{N_\ell,M}(\rho,\phi),
\] 
where \(Q_{N_\ell,M}\) describes the Landau-type states in the \(\rho\)–\(\phi\) plane, while \(f_{N_z,P}(z)\)   governs the axial motion in an effective 1D Coulomb-type potential. This simplification, which was also pursued in \cite{Sch39}, greatly reduces the complexity of solving the full three-dimensional Schrödinger equation.

In the perpendicular plane, the electron's dynamics is governed by the Magnetic field \(\mathbf B=B\hat z\),
\begin{equation}
\hat H_\perp=\frac{1}{2m}\Bigl(\hat{\mathbf p}+e\mathbf A\Bigr)^2,\qquad
\mathbf\nabla\times\mathbf A=B\hat z .
\end{equation}
where we choose the symmetric gauge \(\mathbf A=\tfrac12\,\mathbf B\times\mathbf r=\tfrac{B}{2}(-y,\,x,\,0)\). Introducing the cyclotron radius  
\(\displaystyle r_c=\sqrt{{\hbar}/{|e| B}}\)
and the kinetic momenta
\(\hat{\pi}=\hat{\mathbf p}+e\mathbf A\), two independent oscillator pairs are defined by \cite{Ton16,Aro24}:
\begin{eqnarray}
\begin{aligned}
a&=\frac{r_c}{\sqrt{2}\hbar}\bigl(\pi_x-i\pi_y\bigr)\\
b&=\frac{1}{\sqrt{2}\,r_c}\bigl(x+iy\bigr)+
\frac{r_c}{\sqrt{2}\hbar}\bigl(\pi_x+i\pi_y\bigr),
\end{aligned}
\label{Eq_ladder}
\end{eqnarray}
where they hold the commutation relations
\[
[a,a^\dagger]=[b,b^\dagger]=1,\qquad [a,b]=[a,b^\dagger]=0 .
\]
The  Landau level index (cyclotron index) \(N_\ell\) is changed by \(a,a^\dagger\), while \(b,\,b^\dagger\) move the guiding centre index $N_b$. Since the angular momentum is $L_z=\hbar(b^\dagger b-a^\dagger a)$, the magnetic quantum number $M=N_b-N_a$ would be changed by both ladder operators.  The guiding center ladder operators commute with $H_\perp$
\begin{equation}
\hat H_\perp=\hbar\omega_c\Bigl(a^\dagger a+\tfrac12\Bigr),\qquad
\omega_c=\frac{ |e| B}{m}.
\label{Eq_HPerp}
\end{equation}
and hence costs no energy.

The simultaneous vacuum of both oscillators  (\(a|0,0\rangle=b|0,0\rangle=0\)) can be written in real space as
\[
\langle\mathbf r|0,0\rangle
     =\frac{1}{\sqrt{2\pi}\,r_c}\;
       \exp\!\bigl(-\rho^2/4\,r_c^{2}\bigr)\quad
       (\rho=\sqrt{x^2+y^2}),
\]
i.e., a Gaussian centred on the origin.

\begin{figure*}
\centering
\includegraphics[trim=18 290 67 0, clip, width=\linewidth]{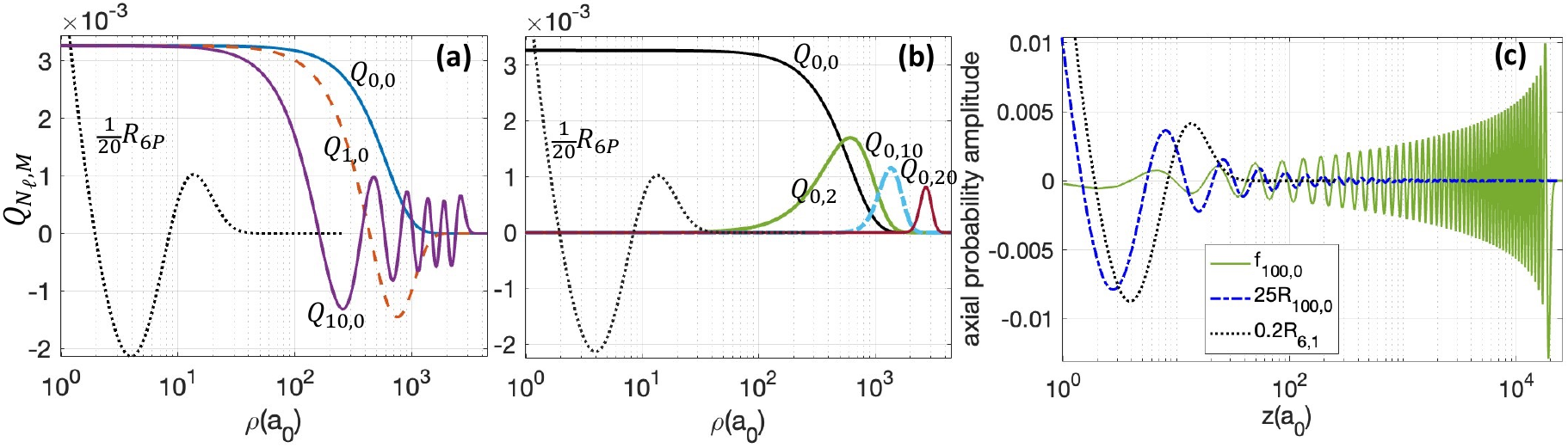}
\caption{
{\bf Wavefunctions of rLandau States. (a,b) } Transverse probability amplitude distribution \(Q_{N_\ell,M}(\rho,\phi)\), and {\bf (c)} axial profile \(f_{N_z}(z)\).  
{\bf (a)} For $M = 0$, the rLandau wavefunction is strongly localized near the ionic core, resulting in significant overlap with the $6P$ state and a large dipole transition amplitude.
{\bf (b)} For \(M \neq 0\), the wavefunction has negligible amplitude at the core, leading to over three orders of magnitude suppression in transition strength.  
   In the case of $N_\ell = 0$, $M \neq 0$, the radial probability distribution exhibits a single peak at $\rho = R_c \sqrt{2|M| + 1}$, forming a characteristic ring-shaped electron cloud associated with circular states.
{\bf (c)}Comparison between the axial rLandau wavefunction $f(z)$ and the 3D hydrogenic radial wavefunction $R_{nl}(r)$. Magnetic confinement squeezes the transverse distribution, effectively modifying the Coulomb potential along the axial direction. As a result, $f(z)$ extends significantly farther from the core compared to the rapidly decaying hydrogenic function $R_{nl}(r)$ under a $1/r$ potential. 
 This spatial separation leads to a strongly suppressed overlap with inner hydrogenic states—illustrated by the dotted line showing $R_{6,1}(r)$—and thereby substantially enhances the lifetime of the rLandau state.}\label{Fig_OverlapWF}
\end{figure*}

The Landau eigenstates are then obtained by applying the creation operators
\[
\lvert N_\ell, M\rangle
=\frac{(a^\dagger)^{N_\ell}\,(b^\dagger)^{N_\ell+M}}{\sqrt{N_\ell!\,(N_\ell+M)!}}\lvert0,0\rangle,
\]
where the power of $b^{\dagger}$ must be positive and hence $M>-N_\ell$. Considering the spatial form of the raising operators
\begin{eqnarray}
b^\dagger\;=\;\frac{\rho\,e^{i\varphi}}{\sqrt2\,r_c}
           -\frac{r_c\,e^{i\varphi}}{\sqrt2}\bigl(\partial_\rho+\tfrac{i}{\rho}\partial_\varphi\bigr)\\ \nonumber
 a^\dagger=-\tfrac{r_c\,e^{-i\varphi}}{\sqrt2}(\partial_\rho-\tfrac{i}{\rho}\partial_\varphi)+\tfrac{\rho\,e^{-i\varphi}}{\sqrt2\,r_c}
\end{eqnarray}
  and applying them to the Gaussian vacuum, forms the Landau state $\bra{\bf{r}}N_\ell,M\rangle$. Note that  \((a^\dagger)^{N_\ell}\) converts the Gaussian into a Laguerre polynomial of order \(N_\ell\).  The resulting Landau function for $M>-N_\ell$ would be      
\begin{eqnarray}\label{Eq_LandauWF}
&&Q_{N_\ell,M}(\rho,\phi)=\frac{e^{iM\phi}}{\sqrt{2\pi}r_c}
\sqrt{\frac{(N_\ell-\frac{|M|-M}{2})!}{(N_\ell+\frac{|M|+M}{2})!}}\,\\ \nonumber
&&
e^{-\rho^{2}/4r_c^{2}}\,\Bigl(\tfrac{\rho^2}{2r_c^2}\Bigr)^{|M|/2}\, \mathcal{L}_{N_\ell-\frac{|M|-M}{2}}^{|M|}\!\Bigl(\tfrac{\rho^{2}}{2r_c^{2}}\Bigr),\qquad
\end{eqnarray}
where \(\mathcal{L}\) is the associated Laguerre polynomial. 
Considering the first two terms, for $M\neq 0$, the biggest peak forms at a position defined by the cyclotron radial distance $r_c$ and $|M|$ while for $M=0$ it is concentrated at the origin, see Fig.~\ref{Fig_OverlapWF}(a-b). 

The second term shifts the states with larger $|M|$ away from the center. For example, for $N_\ell=0$ the position of the single peak is given by $r_c\sqrt{2|M|+1}$. 
The number of radial nodes  $N_\rho=N_{\ell}-\frac{|M|-M}{2}$ and radial anti-nodes  $N_{\rho}+1$ are determined by the associated Laguerre polynomial. 
Considering Eq.~\ref{Eq_HPerp}, the Landau eigenenergies are given by
$  \hbar\omega_c(N_{\ell}+\frac{1}{2})$,
where $\omega_c= \frac{|e|B}{m}$ is the cyclotron frequency. For $B=2.5$T, the cyclotron frequency is $\omega_c/2\pi=70$GHz.
Note that the quadratic magnetic potential is the dominant term in Eq.~\ref{Eq_H} for all $N_\ell$ and $M$ quantum numbers when averaged over the Landau wave-function of Eq.~\ref{Eq_LandauWF}.

  Even though the Coulomb interaction is spherically symmetric if the electron wavefunction is forced into a cylinder of small radius \(\rho \lesssim \sqrt{\langle\rho^2\rangle}\propto r_c\) by the strong magnetic field, the wavefunction’s extension along the \(z\) axis can still be large, resembling a quasi-1D problem. In such a scenario, the perpendicular coordinate will be Landau-like states, and the residual motion along \(z\) can be viewed almost independently, albeit with a modified effective potential that accounts for the electron’s transverse distribution.
The equation governing motion along \( z \)-axis is:
\begin{equation}\label{Eq_Vz}
   \left[-\frac{\hbar^2}{2m}\frac{d^2}{dz^2} +V_{N_\ell M}(z)\right]f(z) = E_z^{N_zN_\ell M}f(z),
\end{equation}
with the effective potential given by integrating the Coulomb term over perpendicular coordinates:
\begin{equation}\label{Eq_Vz}
V_{N_\ell M}(z) =-\iint |Q_{N_\ell,M}(\rho,\phi)|^2 \frac{e^2}{4\pi\epsilon_0 \sqrt{\rho^2+z^2}} \rho d\rho d\phi.
\end{equation}
In the regime where the axial extension of the electron wavefunction is much larger than its radial spread, the effective potential approximates a shifted Coulomb form \( V(z) \approx -e^2/[4\pi\epsilon_0(|z|+d)] \), with values for the shift \( d \) provided numerically (see Table \ref{tab_d}).

 A physical solution requires both the wave function and its slope to be continuous at z = 0. The matching conditions at that point lead to an eigenvalue correction $E_z^{N_zPN_\ell M}=-Ry/(N+\delta^{N_\ell M}_{even/odd})^2$  which is satisfied by introducing the quantum defect  $\delta^{N_\ell M}_{even/odd}$, see Table \ref{tab_d}. The total energy of the Rydberg-Landau level is then given by 
\begin{equation}\label{Eq_Energy}
E_{N_zPN_\ell M}=\hbar \omega_c(N_\ell+1/2)-\frac{Ry}{(N_z+\delta^{N_\ell M}_{P})^2}.
\end{equation} 
where $P=0,1$ corresponds to even and odd parity of $f(z)$.  The lowest $N_\ell=0$ branch remains bound for all $N_z$. At 2.5T for $N_z=100$ (80) the states with $N_\ell<5$ ($<7$) are bound to the nucleus.

Solutions \( f(z) \) are numerically obtained using the Numerov method. Fig.~\ref{Fig_OverlapWF}c compares the 1D axial rLandau wavefunction $f(z)$ in the presence of a magnetic field with the radial wavefunction \(R_{nl}(r)\) in the absence of a magnetic field. Unlike the standard 3D hydrogen-atom potential \(V(r)\propto 1/r\), the tight transverse confinement by magnetic field modifies the effective Coulomb interaction along \(z\) in rLandau states. As a result, the 1D wavefunction can exhibit substantial amplitude far from the ionic core, reflecting the altered balance between kinetic and potential energy in this lower-dimensional setting.
As an additional check on the axial and radial extent, Fig.~\ref{Fig_OverlapWF}c for $N_z = 100$ gives $\langle |z| \rangle / r_c \approx 24$, corroborating the assumption stated above.

By contrast, in the usual 3D hydrogen atom, the radial wavefunction \(R_{nl}(r)\) quickly decays at large \(r\), partly because the electron spreads out in all three dimensions and is subject to the full \(1/r\) Coulomb potential. Normalization constraints also differ between 1D and 3D problems. In 3D, the wavefunction must vanish sufficiently fast at large \(r\) to avoid divergent volume integrals, whereas in 1D, with a cylindrical wavefunction that is squeezed in transverse dimension, even some degree of increase before final termination can still satisfy normalization. Hence, although both systems feature Coulomb-like potentials, the geometry imposed by the strong magnetic field in the rLandau scenario leads to distinctly different Schrodinger equation and subsequently different wavefunction behaviors along the \(z\) axis compared to a standard 3D hydrogenic orbital.

\begin{table}[h!]
\begin{center}
\begin{tabular}{||c|c|c|c||c|c|c|c||}
    \hline
    $\mathbf{Q_{N_\ell,M}}$ & $\mathbf{d(a_0)}$ & $\boldsymbol{\delta_{odd}}$ & $\boldsymbol{\delta_{even}}$ & $\mathbf{Q_{N_\ell,M}}$ & $\mathbf{d(a_0)}$ & $\boldsymbol{\delta_{odd}}$ & $\boldsymbol{\delta_{even}}$ \\
    \hline
    $Q_{0,0}$ & $0.104$ & $0.61$ & $0.61$ & $Q_{3,2}$ & $355$ & $0.07$ & $0.12$ \\
    $Q_{0,1}$ & $178$ & $0.25$ & $0.25$ & $Q_{3,6}$ & $452$ & $0.83$ & $0.35$ \\
    $Q_{0,2}$ & $227$ & $0.78$ & $0.78$ & $Q_{4,-4}$ & $299$ & $0.27$ & $0.84$ \\
    $Q_{0,6}$ & $358$ & $0.24$ & $0.24$ & $Q_{4,-3}$ & $308$ & $0.48$ & $0.12$ \\
    $Q_{1,-1}$ & $178$ & $0.25$ & $0.25$ & $Q_{4,-2}$ & $317$ & $0.74$ & $0.3$ \\
    $Q_{1,0}$ & $0.3$ & $0.73$ & $0.73$ & $Q_{4,-1}$ & $327$ & $0.03$ & $0.47$ \\
    $Q_{1,1}$ & $238$ & $0.1$ & $0.1$ & $Q_{4,0}$ & $1.2$ & $0.68$ & $0.31$ \\
    $Q_{1,2}$ & $275$ & $0.65$ & $0.05$ & $Q_{4,1}$ & $364$ & $0.89$ & $0.37$ \\
    $Q_{1,6}$ & $392$ & $0.03$ & $0.03$ & $Q_{4,2}$ & $390$ & $0.48$ & $0.07$ \\
    $Q_{2,-2}$ & $226$ & $0.25$ & $0.75$ & $Q_{4,6}$ & $481$ & $0.4$ & $0.87$ \\
    $Q_{2,-1}$ & $237$ & $0.58$ & $0.2$ & $Q_{5,-5}$ & $330$ & $0.05$ & $0.6$ \\
    $Q_{2,0}$ & $0.7$ & $0.56$ & $0.17$ & $Q_{5,-4}$ & $338$ & $0.25$ & $0.18$ \\
    $Q_{2,1}$ & $285$ & $0.9$ & $0.26$ & $Q_{5,-3}$ & $346$ & $0.46$ & $0.04$ \\
    $Q_{2,2}$ & $317$ & $0.75$ & $0.16$ & $Q_{5,-2}$ & $355$ & $0.67$ & $0.22$ \\
    $Q_{2,6}$ & $423$ & $0.25$ & $0.64$ & $Q_{5,-1}$ & $364$ & $0.87$ & $0.41$ \\
    $Q_{3,-3}$ & $265$ & $0.38$ & $0.81$ & $Q_{5,0}$ & $1.7$ & $0.96$ & $0.47$ \\
    $Q_{3,-2}$ & $275$ & $0.65$ & $0.05$ & $Q_{5,1}$ & $399$ & $0.66$ & $0.24$ \\
    $Q_{3,-1}$ & $285$ & $0.89$ & $0.25$ & $Q_{5,2}$ & $422$ & $0.2$ & $0.69$ \\
    $Q_{3,0}$ & $1$ & $0.48$ & $0.3$ & $Q_{5,3}$ & $445$ & $0.67$ & $0.15$ \\
    $Q_{3,1}$ & $327$ & $0.4$ & $0.45$ & $Q_{5,6}$ & $508$ & $0.03$ & $0.5$ \\
    \hline
\end{tabular}
\end{center}
\caption{For large \(N_z\), where the axial extension exceeds the radial width, the effective 1D Coulomb potential simplifies to 
\(\displaystyle V(z) = -\frac{e^2}{4\pi\epsilon_0\,[|z| + d]}\). 
This table lists the numerically extracted shift \(d\), along with the quantum defects \(\delta_{\rm even/odd}\) that ensure continuity of the wavefunction and its derivative at \(z = 0\). }\label{tab_d}
\end{table}

\section{Selection Rules of Ryd-Landau Transitions}

To excite the Rydberg–Landau (rLandau) states in \(^{87}\mathrm{Rb}\), we employ a two-photon excitation scheme via the intermediate state \(6P_{1/2}\). The selection rules governing these transitions are determined primarily by angular momentum and parity considerations.

For dipole transitions between the intermediate state \(6P\) and an rLandau state \(\ket{N_z,P,N_\ell,M}\), the magnetic quantum number \(M\) remains a good quantum number due to cylindrical symmetry around the magnetic field axis. The corresponding transition dipole moment, proportional to \(\bra{N_z,P,N_\ell,M}r_q\ket{6P,m}\) (with polarization \(q=0,\pm 1\) for $r_0=z$, $r_{\pm}=(x\pm iy)/\sqrt{2}$), includes the integral \(\int e^{i(m+q-M)\phi} d\phi\). This integral vanishes unless:
\begin{equation}
    M - m = q = 0, \pm 1.
\end{equation}
This expresses the selection rule for changes in the magnetic quantum number, dependent on the polarization of the excitation laser.

    \begin{figure}
\centering
    \includegraphics[trim=0 270 375 0, clip, width=\linewidth]{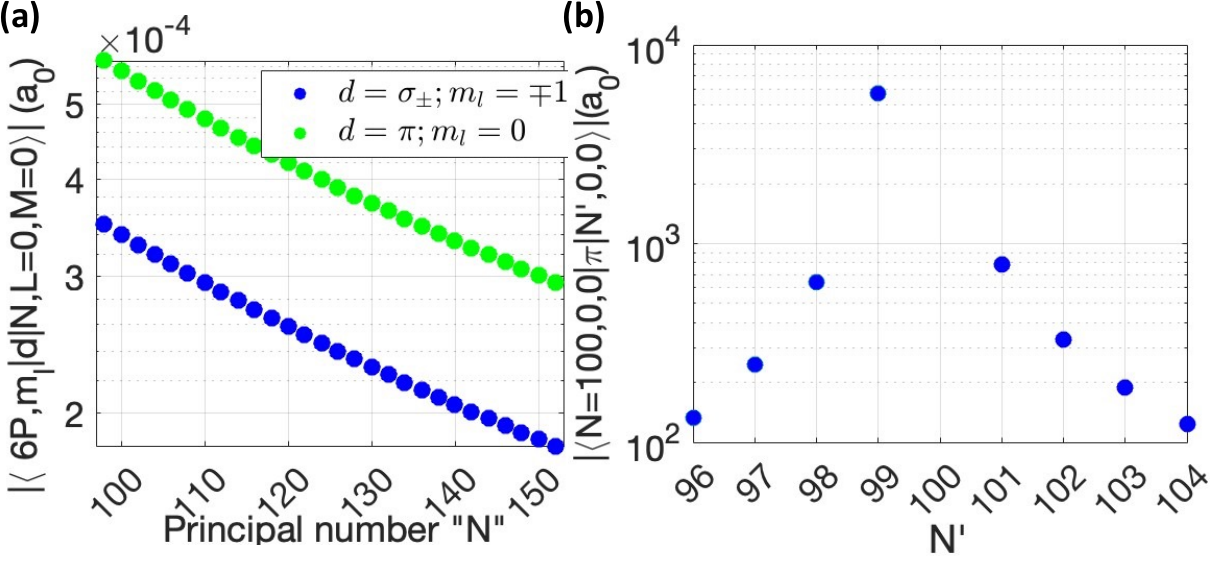}
\caption{{\bf Scaling of the transition dipoles} for (a)  ground to rLandau states $\langle 6P,m_l |d|N_z',N_\ell'=0,M'=0\rangle$ and (b) inter rLandau states.} \label{Fig_DRate}
\end{figure}

Parity imposes an additional constraint. Under spatial inversion \(\mathbf{r}\rightarrow -\mathbf{r}\), the parity of a hydrogen-like orbital scales as \((-1)^l\), while the parity of the rLandau wavefunction is determined by the combined axial parity \(P\) and the magnetic quantum number \(M\), expressed as \((-1)^{P+M}\). Since electric dipole transitions change parity, initial and final states must have opposite parities. Thus, the parity selection rule for transitions between the hydrogenic and rLandau states becomes:
\begin{equation}
    P + M + l = 2k + 1, \quad k \in \mathbb{Z}.
\end{equation}
These selection rules are summarized in Table~\ref{Tab_SelectionRules}.

For transitions between rLandau states, similar considerations apply. Under linearly polarized (\(\pi\)) light, the transition dipole moment is factorized as:
\begin{equation}
    d_{\pi} = \bra{f_{N_{z1},P_1}} z \ket{f_{N_{z2},P_2}} \bra{Q_{N_{\ell1},M_1}} Q_{N_{\ell2},M_2}\rangle.
\end{equation}
The integral over the axial \(z\)-direction mandates a change in parity \(P\) and principal quantum number \(N_z\), whereas the Landau quantum numbers \(N_\ell\) and \(M\) remain unchanged due to orthogonality: \(\bra{Q_{N_{\ell1},M_1}}Q_{N_{\ell2},M_2}\rangle = \delta_{N_{\ell1},N_{\ell2}}\delta_{M_1,M_2}\).

For right- or left-circular polarization (\(\sigma_{\pm}\)), the transition dipole moment is:
\begin{equation}
    d_{\sigma_{\pm}} = \bra{f_{N_{z1},P_1}} f_{N_{z2},P_2}\rangle \bra{Q_{N_{\ell1},M_1}} x \pm i y \ket{Q_{N_{\ell2},M_2}}.
\end{equation}
The axial integral preserves both parity \(P\) and principal number \(N_z\), as the functions \(f_{N_z}\) are orthogonal for different \(N_z\), and integration over odd-parity integrands vanishes. 
However, integration in the transverse plane allows a change in the Landau quantum numbers.
The circular transition dipole moments can be writen in terms of the ladder operators (see Eq.~\ref{Eq_ladder})  
\begin{equation}
\frac{x+ i y}{\sqrt2} =\sqrt2\,r_c\bigl(b^\dagger - a\bigr),\quad
\frac{x-iy}{\sqrt2} =\sqrt2\,r_c\bigl(b - a^\dagger\bigr).
\end{equation}
The corresponding transition would be
\begin{eqnarray}
  \label{Eq_sigmapmSelectRule}
&&\hat \sigma_+\lvert N_\ell,M\rangle=\sqrt{2}r_c\\ \nonumber
&& \Bigl( \sqrt{N_\ell+M+1}\lvert N_\ell,M+1\rangle - \sqrt{N_\ell}\,\lvert N_\ell-1,\,M+1\rangle \Bigr) \\ \nonumber
&&\hat \sigma_-\lvert N_\ell,M\rangle=\sqrt{2}r_c\\ \nonumber
&&
  \Bigl(
    \sqrt{N_\ell+M}\lvert N_\ell,M-1\rangle -\sqrt{N_\ell+1}\lvert N_\ell+1,M-1\rangle
  \Bigr)
\end{eqnarray}
All detailed selection rules for these transitions are presented succinctly in Table~\ref{Tab_SelectionRules}, and confirmed numerically.

\begin{table}[h]
\begin{center}
\begin{tabular}{|c|c|c|}
\hline
  \text{transition}&  Pol &  \text{Selection Rules}   \\
    \hline
    \hline
    & &      \\
  & $\pi$   & $M=m$; \, $m+l+P=2k-1$  \\
    & &      \\
  $\text{Hyd}$ $\rightarrow$ $\text{rLandau}$    & $\sigma_{+}$ & $M=m+1$; \, $m+l+P=2k$   \\
    & &      \\
   $\ket{nlm}\bra{N_zPN_\ell M}$ &  $\sigma_{-}$  &$M=m-1$; \,  $m+l+P=2k$   \\
    & &      \\
    \hline
   & &      \\
 \multirow{3}{*}{\rotatebox{40}{$\text{rLandau} \rightarrow \text{rLandau}$}}  & $\pi$ & $\Delta$M=0; \,  $\Delta$P=1; \,  $\Delta N_\ell=0$   \\
     & &      \\
 & $\sigma_{+}$ & $\Delta$M=+1; $\Delta$(P,N$_z$)=0; $\Delta N_\ell=0,-1$   \\
    & &      \\
    & $\sigma_{-}$  & $\Delta$M=-1; $\Delta$(P,N$_z$)=0; $\Delta N_\ell=0,+1$   \\
    & &      \\
    \hline
\end{tabular}
\end{center}
\label{Tab_SelectionRules}
\caption{Selection rules in transition from hydrogenic to rLandau states, as well as the transition among the rLandau states as a function of the laser polarization.} \label{Tab_SelectionRules}
\end{table}

\section{Excitation of Ryd-Landau States}

\begin{figure*}[ht]
    \centering
    \includegraphics[trim=94 308 53 0, clip, width=\linewidth]{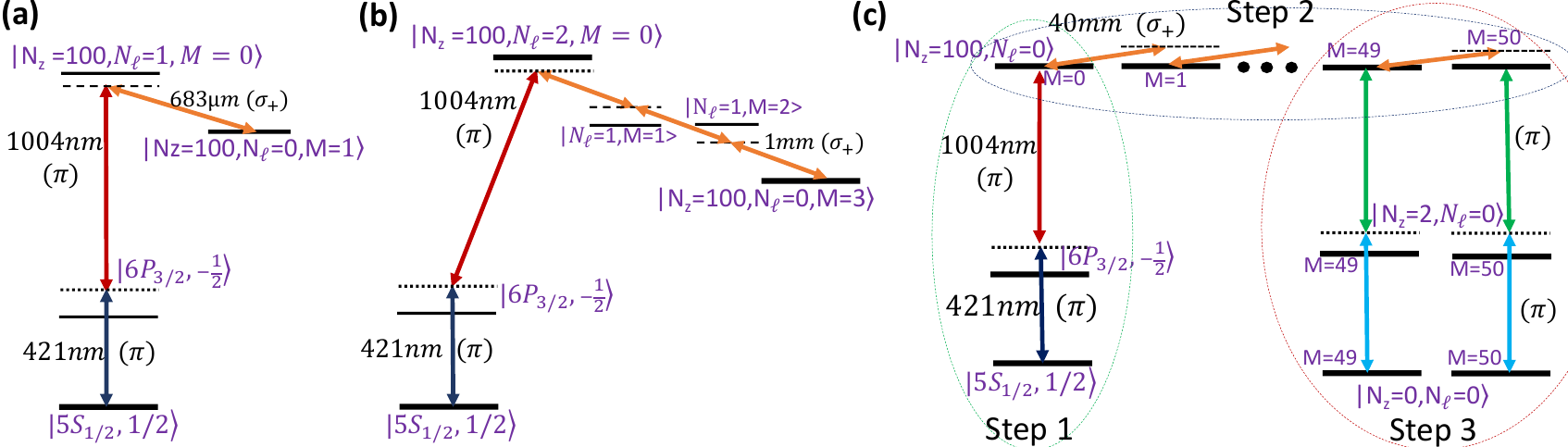}
\caption{ {\bf Level scheme for the excitation of rLandau states.} A two-photon transition via the intermediate hydrogenic $6P$ state is used to excite the $\protect \ket{N_z, N_\ell, M = 0}$ state. While the $M = 0$ state exhibits strong dipole coupling to the Coulombic ground state—facilitating efficient excitation—subsequent transitions to $M \neq 0$ states reduce the wavefunction overlap with Coulombic eigenstates, thereby significantly enhancing the lifetime. (a) The $\protect \ket{N_z = 100, N_\ell = 0, M = 1}$ state can be reached via a three-photon process, with intermediate states including the hydrogenic $6P$ level and the rLandau state $\protect \ket{N_z = 100, N_\ell = 1, M = 0}$. (b) Higher magnetic quantum number states such as $\protect \ket{N_\ell = 0, M = 3}$ can be accessed by appropriately tuning the laser frequencies in a similar multi-photon excitation scheme. (c) An alternative approach involves exciting a coherent superposition of magnetic sublevels, as described by Eq.~\ref{Eq_state}, after steps 1 \& 2 marked in the figure. The detunings in this process are compensated by sufficiently strong Rabi frequencies. Step 2 could be reversed to retrieve the $\protect \ket{N_\ell = 0, M = 0}$ state by reversing the polarization of the same field.  Final de-excitation to the $\protect \ket{N_z = 0, N_\ell = 0, M}$ state, depicted in step 3, enables preparation of ultra-long-lived circular rLandau states.}\label{Fig_LevelScheme}
\end{figure*}

To experimentally excite rLandau states, we employ a two-photon transition via the intermediate state \(\ket{6P}\). Figure~\ref{Fig_DRate}a illustrates the magnitude of the transition dipole moment \( d_{6P}^{N_zN_\ell M} = \bra{N_z,N_\ell,M} r_q \ket{6P} \) as a function of the quantum numbers \(N_z\). The transition dipole scaling can be understood by examining the spatial overlap between the \(6P\) state and the rLandau wavefunctions, both along and perpendicular to the magnetic field direction, see Fig.~\ref{Fig_OverlapWF}.
In transitions involving the \(6P\) intermediate state, 
The spin-orbit coupling for the low-lying Coulombic state must be taken into account.  
For example, exciting the rLandau state \(\ket{N_z,P=0,N_\ell=0,M=0}\) with $\pi$ polarized light from \(\ket{6P_{3/2}, m_j=-1/2} = R_{6p}(\sqrt{1/2}\,Y_{10}\chi_{\downarrow}+\sqrt{1/2}\,Y_{1,-1}\chi_{\uparrow})\) involves only the \(Y_{10}\) component due to selection rules, introducing an additional Clebsch–Gordan factor of \(\sqrt{1/2}\) to the transition dipole. 

Landau states with magnetic quantum number \(M=0\) exhibit non-zero probability density at \(\rho=0\), enhancing their overlap with the intermediate \(6P\) state and thus increasing the transition dipole moment \( d_{6P}^{N_zN_\ell M} \), as indicated in Eq.~\eqref{Eq_LandauWF} and shown in Fig.~\ref{Fig_OverlapWF}a. 
 For \(M=0\) states, the dipole transition moment is nearly independent of the Landau level index \(N_\ell\). 
 Conversely, the transition dipole to states with \(M\neq0\) is more than three orders of magnitude smaller, due to the absence of population close to the ionic core. 

While the $M = 0$ state possesses a strong transition dipole moment that enables efficient laser excitation, states with $M \neq 0$ benefit from significantly enhanced lifetimes due to reduced overlap with Coulombic ground states. Two general strategies can be employed to access $M\neq 0$ long-lived states. The first involves selectively exciting a single magnetic quantum number, while the second targets a superposition of multiple $M$ states. In the first approach, illustrated in Fig.~\ref{Fig_LevelScheme}a, the state $\ket{N_z = 100, N_\ell = 0, M = 1}$ is populated via a three-photon excitation pathway involving off-resonant intermediate states, namely the $6P$ level and $\ket{N_z = 100, N_\ell = 1, M = 0}$. Accessing higher $M$ states requires careful tuning of the excitation frequencies. For instance, Fig.~\ref{Fig_LevelScheme}b depicts a scheme designed to excite the state $\ket{N_\ell = 0, M = 3}$, showcasing the flexibility of this method in engineering target magnetic quantum numbers.

The second approach, shown in Fig.~\ref{Fig_LevelScheme}c, begins from the state $\ket{N_z = 100, N_\ell = 0, M = 0}$ and employs a right-circularly polarized electromagnetic wave with wavelength $\lambda \geq 40$ mm. The applied radiation drives a cascade of detuned transitions that sequentially increase the magnetic quantum number. When the Rabi frequency $\Omega$ is large enough that laser power broadening exceeds the detuning, the system evolves into a coherent superposition of magnetic sublevels:
\begin{equation}
\ket{\psi(t)} = \sum_{M = 0}^{\infty} \frac{(-i \Omega t)^M}{\sqrt{M! \, \exp[(\Omega t)^2]}} \ket{N_\ell = 0, M}.
\label{Eq_state}
\end{equation}
To return the system to its initial state $\ket{N_z, N_\ell = 0, M = 0}$, the polarization is simply reversed to left-circular for an equivalent duration. For this purpose, we employ a tunable microwave horn, similar to those used in previous experiments involving coherent microwave and millimeter-wave transitions between Rydberg states \cite{Bai24, Coh21, Bai19}.

Additionally, another class of ultra-long-lived states can be created by driving transitions from \(\ket{N_z=100,P=0,N_\ell=0,M}\) to the ground-\(N_z\) state \(\ket{N_z=0,P=0,N_\ell=0,M}\) using two-photon transition via the intermediate state \(\ket{N_z=2,P=1,N_\ell=0,M}\), see the step 3 in Fig.~\ref{Fig_LevelScheme}c. This target state lacks any dipole-allowed decay paths. Notably, for \(M\neq0\), the electron distribution takes on a doughnut-like shape resembling Coulombic circular states (Fig.~\ref{Fig_OverlapWF}d), characterized by a radius \(\rho_d=R_c\sqrt{|M|+1}\). Unlike Coulombic circular states, whose coherent excitation remains experimentally challenging \cite{Car20, Sig17}, these long-lived rLandau circular states can be coherently prepared using standard multi-photon dipole transitions, making them highly suitable for quantum technology applications.

 \begin{figure*}
\centering
    \includegraphics[trim=0 335 0 0, clip, width=\linewidth]{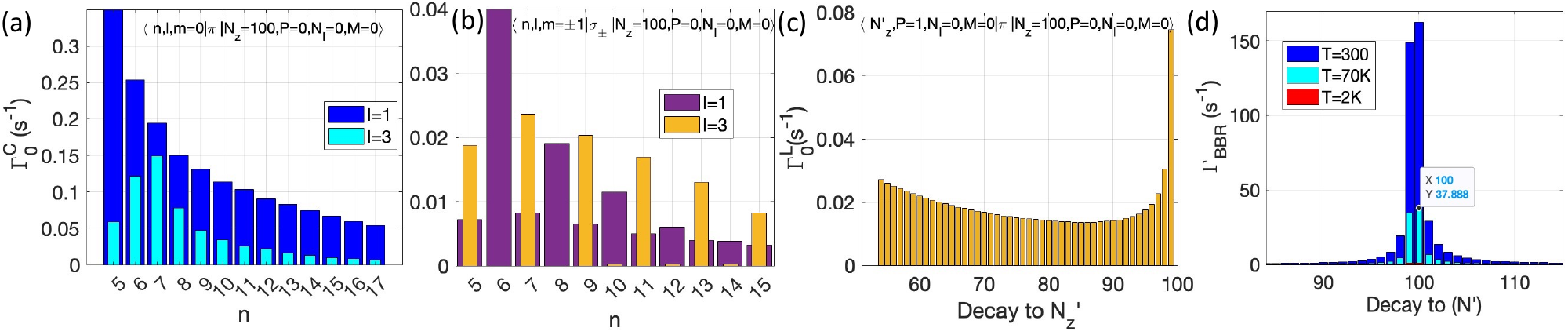}
\caption{{\bf The error budgeting} of the spontaneous decay rates of the even parity Landau state $|N_z=100, P=0, N_\ell=0, M=0\rangle$  to the lower (a,b) hydrogenic and (c) Landau states as well as (d) Black-Body radiation (BBR) induced depopulation to neighbouring rLandau states. The emitted photon is (a,c) linearly and  (b) circularly polarized. The decay rate to $l>3$ angular momentum numbers is more than five orders of magnitude smaller. The Rydberg-Landau state with $N_\ell=0$ has the lowest Landau energy and hence does not decay to other $N_\ell$ states. It only decays to lower $N_z$ quantum numbers and emits linear light with the rates presented in (c). }\label{Fig_ACoef}
\end{figure*}


\section{lifetime of Ryd-Landau states}
The lifetime of the rLandau states is defined by the spontaneous emission and black-body radiation depopulations.
The rate of spontaneous emission from state $i$ to $f$ is evaluated by the Einstein coefficient:
\begin{equation}
    A = \frac{4e^2}{3\hbar}(\frac{\omega_{fi}}{c})^3|r_{fi}|^2
    \label{e41}
\end{equation}
where $r_{fi}=\langle f|r|i \rangle$ is the dipole transition and  $\omega_{fi}=\frac{E_f-E_i}{\hbar}$ is the transition frequency.
The total rate of spontaneous transitions from the rLandau state to all lower-lying states would be given by
\begin{equation}
 \Gamma_0 = \sum_{E_f<E_i} A(i \rightarrow f). 
 \label{e42}
\end{equation}

The rLandau state $\ket{N_z, P, N_\ell=0, M=0}$ with even/odd parity, emits a $\sigma_{\pm}$ light and decays to a hydrogenic state $\ket{n,l,m_l=\pm1}$ with odd/even magnetic quantum numbers $l$ respectively, see table \ref{Tab_SelectionRules}.
Note that the decay to states with larger $l$ is weaker and we can neglect the decay to states with $l>3$ as they are more than five orders of magnitude smaller than the decay to $l=0$ or $l=1$. Fig.~\ref{Fig_ACoef} shows the decay rate of the initial state $\ket{100, P=0, N_\ell=0, M=0}$ to the hydrogenic state as well as the lower Landau states.
The Rydberg-Landau state with $N_\ell=0$ has the lowest Landau energy and hence does not decay to other $N_\ell$ states. It only decays to lower $N_z$ quantum numbers and emits linear light with the rates presented in Fig.~\ref{Fig_ACoef}(c).

The total rate of black-body radiation (BBR) induced depopulation can be written in a similar form, taking into account transitions to both lower and higher states:
\begin{equation}\label{e41}
    \Gamma_{BBR} = \sum_{f} A(i \rightarrow f) \frac{1}{\exp(\omega_{if}/kT)-1}
\end{equation}
where $k$ is the Boltzman constant and $T$ is the temperature. Finally, the effective lifetime of the Rydberg-Landau state is determined by the inverse sum of spontaneous and BBR-induced depopulation rates $\tau=(\Gamma_0+\Gamma_{BBR})^{-1}$. The lifetime of a few rLandau states is presented in Table \ref{Tab_Lifetime}.

\section{Ultra-Long-Lived Ryd-Landau States}

We introduce rLandau states characterized by exceptionally long lifetimes, attributed primarily to three distinct factors: significantly reduced overlap with low-lying Coulombic states, restricted dipole-allowed decay channels, and energetic ground-state positioning while maintaining strong dipole-dipole interactions.

Concerning wavefunction overlap, the longitudinal wavefunction $f(z)$ for rLandau states exhibits substantially less overlap with inner Coulombic $P$-orbitals, such as the $5P$ state, compared to traditional Rydberg states. This reduction occurs as the wavefunction is spatially displaced further away from the ionic core, as illustrated in Fig.~\ref{Fig_OverlapWF}(c). Furthermore, transitions between states with different magnetic quantum numbers $M$ significantly alter this overlap. Specifically, initial excitation can efficiently occur via states with $M=0$, followed by prolonged storage in states with $M\neq 0$ due to their minimal overlap with low-lying Coulombic states, see Fig.~\ref{Fig_OverlapWF}(a,b).

\begin{table}[h]
\begin{center}
\begin{tabular}{||c||c|c||c|c|c||c|c|c||}
\hline
  $|N_z,N_\ell,M\rangle$ & $ \Gamma_0^C$ &$ \Gamma_0^L$ & \multicolumn{3}{c||}{$\Gamma_{BBR}$(s$^{-1}$) } &  \multicolumn{3}{c||}{ \text{Lifetime}(ms)}\\
 &(s$^{-1}$)& (s$^{-1}$)& \multicolumn{3}{c||}{} &  \multicolumn{3}{c||}{}\\
     \hline
  & & & 300K & 70K& 2K&300K& 70K & 2K \\
    \hline
   $|100,0,0\rangle$ & 2.5& 0.87&207  &48 & 0.85& 4.8& 19.5&237 \\
  $|$100,0,M$\neq$0$\rangle$ & $<$10$^{-3}$& 0.87 & 207 & 48& 0.85& 4.8& 20.2& 383\\
  $|0,0,M \rangle$ & 0 &0 & 41 & 7.8& 1.1& 24 & 128 &910  \\
  $|0,0,0 \rangle$ & 0 &0 &$<$10$^{-10}$& $\approx$0 &$\approx$0& $\infty$ & $\infty$ &$\infty$  \\
  \hline
    Hydrogenic  & & &  & & &  &  &  \\
  \hline
  $|100S_{1/2}\rangle$ & 853 &- &2200& 512& 12& 0.05 &  0.7&  1 \\
    \hline
\end{tabular}
\end{center}
    \caption{{\bf Lifetime of rLandau states.}  
Total decoherence rates are shown, including contributions from spontaneous emission to lower rLandau states (\(\Gamma^L_0\)), coulombic states (\(\Gamma^C_0\)), blackbody radiation-induced depopulation (\(\Gamma_{\mathrm{BBR}}\)), and total lifetimes at various environmental temperatures. 
The rLandau states lifetimes are much larger than the Coulombic Rydberg lifetime presented in the last row.  }\label{Tab_Lifetime}
\end{table}

Regarding dipole-allowed decay channels, selection rules for inter-Landau dipole transitions restrict changes in quantum numbers $N_\ell$ and $M$ exclusively through right- or left-circularly polarized photons. Consequently, an rLandau state $\ket{N_z,N_\ell=0,M}$ is prohibited from spontaneous emission to energetically lower states by emitting circularly polarized photons. The available decay routes are thus limited to transitions to lower-$N_z$ rLandau states $\ket{N_z', N_\ell=0, M}$, with $N_z'<N_z$, via linearly polarized photons.

Another decay pathway involves transitions to hydrogenic states with significantly lower principal quantum numbers. However, these transitions are governed by the selection rule $m \in [M,M\pm1]$. Since transition dipoles to Hydrogenic states with large angular momentum $l>3$ are more than five orders of magnitude smaller than the decay to l=1,3, exciting $M>4$ makes the decay channels to $l<4$ selection rule forbidden.

Finally, rLandau circular states (CS) $\ket{N_z=0, N_\ell=0, M}$ feature ultra-long lifetimes as they represent the energetic ground state of both Coulombic and Landau energies, completely precluding spontaneous emission. This characteristic starkly contrasts with Coulombic CS, which occupies highly excited states where the lifetime is susceptible to distortions. Moreover, rLandau CS demonstrate enhanced robustness against blackbody radiation-induced depopulation since their nearest dipole-allowed transitions are energetically more distant compared to Coulombic analogs. Notably, despite being energetically ground states, rLandau CS possess substantial dipole moments, facilitating strong inter-state interactions.

\section{Interaction of Rydberg–Landau atoms}

The dipole-dipole interaction between two Rydberg–Landau atoms significantly influences their collective dynamics and applications in quantum technology. Consider a pair of identical rLandau states \(\ket{\chi} = \ket{N,L,M}\ket{N,L,M}\). The dipole coupling to another pair of states \(\ket{\psi} = \ket{N_i,L_i,M_i}\ket{N_j,L_j,M_j}\) is described by the dipolar interaction Hamiltonian:
\begin{equation}
V = \frac{e^2}{4\pi\epsilon_0 R^3}\bra{\psi}\mathbf{r}_1\cdot\mathbf{r}_2 - 3(\mathbf{r}_1\cdot\hat{\mathbf{R}})(\mathbf{r}_2\cdot\hat{\mathbf{R}})\ket{\chi},
\end{equation}
where \(\mathbf{r}_{1,2}\) are the position operators of the electrons from the core of each atom, and \(\mathbf{R}\) is the interatomic separation vector.

To clearly illustrate the anisotropy of this interaction, we denote \(\theta\) as the angle between the interatomic vector \(\mathbf{R}\) and the quantization axis (defined by the magnetic field direction). Expressed in spherical tensor components of the dipole operator (\(\sigma_{\pm} = x \pm i y\), \(\pi = z\)), the dipole interaction becomes:
\begin{align}
V_{ij}(R,\theta) &= \frac{e^2}{4\pi\epsilon_0 R^3} 
\bra{\psi_{ij}} \bigg[
    (1 - 3\cos^2\theta)(\sigma_+ \sigma_- + \sigma_- \sigma_+ + \pi\pi) \notag \\
&\quad + 3\sin\theta \cos\theta (\pi\sigma_+ + \pi\sigma_- + \sigma_+\pi + \sigma_-\pi) \notag \\
&\quad + \frac{3}{2} \sin^2\theta (\sigma_+\sigma_+ + \sigma_-\sigma_-) 
\bigg] \ket{\chi} 
\label{Eq_InteractTheta}
\end{align}
In calculating the interaction, the selection rules described in Table~\ref{Tab_SelectionRules} must be considered.

In practice, to facilitate single-site addressing in a quantum processing unit, we arrange the atoms in a two-dimensional optical lattice perpendicular to the magnetic field. In this geometry, the interatomic separation vector is perpendicular to the magnetic field, implying \(\theta = \pi/2\). Under these conditions, the second term in Eq.~\ref{Eq_InteractTheta} vanishes, simplifying the dipole coupling analysis. 

The dipole interaction couples multiple rLandau pair states \(\ket{\psi_{ij}}\), resulting in shifts of the energy levels of the targeted pair state \(\ket{\chi}\). Notably, when the total Landau level index is preserved in a \(\sigma_{\pm}\)-transition  \(2N_\ell = N_{\ell i} + N_{\ell j}\) (which does not change the principal number \(N_z\)), the pair states become degenerate, leading to resonant dipolar interactions.

To explain the off resonant couplings in interaction dynamics, consider two interacting pairs \(\ket{\chi} = \ket{N_z,N_\ell,M}\ket{N_z,N_\ell,M}\) and \(\ket{\psi}_{ij} = \ket{N_{zi},N_{\ell i},M_i}\ket{N_{zj},N_{\ell j},M_j}\) separated in energy by an energy defect \(\delta_{ij} = E(N_{zi},N_{\ell i}) + E(N_{zj},N_{\ell j}) - 2E(N_z,N_\ell)\). The energy shift induced by dipolar coupling between these states can be analyzed using a two-level F\"orster resonance model. In matrix form, the coupled system is described by \cite{Saf10}
\begin{equation}\label{Eq_VanDer}
\begin{pmatrix}
\delta_{ij} & V_{ij}\\
V_{ij}^\dagger & 0
\end{pmatrix}
\begin{pmatrix}
\ket{\psi_{ij}}\\
\ket{\chi}
\end{pmatrix}
=
U_{ij}
\begin{pmatrix}
\ket{\psi_{ij}}\\
\ket{\chi}
\end{pmatrix}.
\end{equation}
where \(V _{ij}= C_{3,ij}/R^3\) defines the dipolar coupling strength. 
Solving the coupled equations, the interaction-induced energy shift \(U(R)\) is found to be:
\begin{equation}\label{Eq_Shift}
U_{ij}(R) = \frac{\delta_{ij}}{2} - \mathrm{sign}(\delta_{ij})\sqrt{\frac{\delta_{ij}^2}{4} + \frac{C_{3,ij}^2}{R^6}},
\end{equation}

The critical distance \(R_c^{ij}\), at which the interaction changes from the resonant dipole–dipole form (\(\sim R^{-3}\)) to the off-resonant van der Waals form (\(\sim R^{-6}\)), is defined by the condition \(\delta_{ij} = C_3/(R_{cr}^{ij})^3\). At large interatomic distances (\(R \gg R_{cr}\)), the interaction reduces to the familiar van der Waals form:
\[
U_{ij}(R) \approx \frac{C_{3,ij}^2}{\delta_{ij} R^6} = \frac{C_{6,ij}}{R^6},
\]
while at short distances (\(R \ll R_{cr}^{ij}\)), the shift approaches the resonant dipole–dipole form:
\[
U_{ij}(R) \approx -\mathrm{sign}(\delta_{ij})\frac{C_{3,ij}}{R^3}.
\]

Summarizing, the total interaction-induced energy shift $U_t=\sum_{i,j}U_{ij}$ experienced by the targeted rLandau pair \(\ket{\chi}\) is given by summing contributions from all coupled pair channels \(\ket{\psi_{ij}}\). Each channel contributes according to its crossover distance \(R_{cr}^{ij}\):
\begin{equation}
U_t(R) = \sum_{i,j}\left(\frac{C_{3,ij}}{R^3}\right)_{R < R_{cr}^{ij}} + \left(\frac{C_{6,ij}}{R^6}\right)_{R > R_{cr}^{ij}}.
\end{equation}



\begin{table}[h!]
\begin{center}
\begin{tabular}{||c||c|c|c|c||}
    \hline   
    $(N_{zi},N_{zj})$ & $C_3$ & $C_6$ & $ \delta$ & $R_{cr}$\\
     &\, (GHz.$\mu$m$^3)$\, & \,(GHz.$\mu$m$^6$) \, & \,(GHz) \, & \,($\mu$m)\, \\
    \hline
    $(99,101)$ & -4.4 & -94 & -0.21 & 2.8  \\    
    $(99,102)$ & 1.8 & 0.54 & 6.2 & 0.67  \\
    $(98,101)$ & -0.48 & -0.03 & -7.2 & 0.41  \\
    $(98,102)$ & -0.2 & -0.05 & -0.81 & 0.6  \\
    $(98,103)$ & -0.12 & -0.0025 & 5.4 & 0.28  \\
    $(97,103)$ & -0.04 & -0.0011 & -1.82 & 0.3  \\  
    \hline
\end{tabular}
\end{center}
\caption{
{\bf Interaction channels for the rLandau pair \(|N_zN_\ell M;N_zN_\ell M\rangle\) under \(\pi\pi\) coupling.}
Shown are the dominant off-resonant dipole-coupled channels for the initial state \(|N_z=100, N_\ell=0, M=0\rangle\). Each entry lists the coupled pair \(|N_{zi}, N_\ell=0, M=0; N_{zj}, N_\ell=0, M=0\rangle\), the energy defect \(\delta\), and the crossover distance \(r_c\) separating the dipolar (\(\propto 1/R^3\)) and van der Waals (\(\propto 1/R^6\)) regimes. See main text for details.} \label{Tab_V}
\end{table}

In analyzing the interaction of rLandau states with large $N_z$, both circular and linear dipole coupling terms play roles.
 Exciting atoms to \(\ket{N_z=100, P=0, N_\ell=0, M=0}\), the only radially coupled pairs are connected by $\sigma_{+}\sigma_{+}$ dipole operators (see Eq.~\ref{Eq_sigmapmSelectRule}), which couples the resonant rLandau pair $\ket{\psi_{ij}}=\ket{N_z=100, N_\ell=0, M=1}\ket{N_z=100,N_\ell=0, M=1}$ with the coupling strength of $C_3=275$MHz.$\mu$m$^3$.  This resonant interaction is ideal for realizing the adiabatic dark state quantum operation schemes \cite{Kha20} where a single resonant coupled pair is required. 
The resonant interaction between \(\ket{N_z=100, P=0, N_\ell=0, M=3}\) excited atoms contains all the combinations of $\sigma_{\pm}$ that do not change the $N_\ell$, adding up to the coupling strength of $C_3=1.1$GHz.$\mu$m$^3$.

The off-resonant $\pi\pi$ coupling conserves the quantum numbers $N_\ell$ and $M$, affecting only the longitudinal quantum number $N_z$ and parity $P$. The dominant contribution to this interaction arises from the transition $\ket{N_z, N_z} \leftrightarrow \ket{N_z-1, N_z+1}$, which simultaneously features the largest dipole matrix element and the smallest energy defect ($\delta$). Detailed properties of these coupled pairs—including their dipole transition strengths and energy defects—are listed in Table~\ref{Tab_V}.

For Rydberg–Landau (rLandau) states with large axial quantum number $N_z$, the axial extent of the wavefunction significantly exceeds its radial confinement, making the $\pi-\pi$ transition the dominant interaction channel, which makes $N_z$ the dominant scaling parameter.
The relevant dipole matrix element $\langle N_z - 1|\pi|N_z\rangle$ scales approximately as $N_z^2$, in contrast to conventional Rydberg–Rydberg interactions, which scale as $N_z^4$. Meanwhile, the energy defect $\delta$ decreases as $N_z^{-3}$, leading to a van der Waals interaction strength that scales as $N_z^7$. For the state $\ket{N_z = 100, N_\ell = 0, M = 0}$, the resulting interaction shift is calculated to be 510~MHz at an interatomic separation of 2~$\mu$m, and 118~MHz at 3~$\mu$m, enabling fast gate operations.

Another particularly compelling regime arises in the interaction between rLandau circular states, such as $\ket{N_z = 0, N_\ell = 0, M}\ket{N_z = 0, N_\ell = 0, M}$. These states can be coherently prepared using the excitation scheme shown in Fig.~\ref{Fig_LevelScheme}. The corresponding electron wavefunction exhibits a characteristic doughnut-shaped probability distribution, with a radius given by $r_c \sqrt{|M| + 1}$, where the cyclotron radius at $B = 2.5\,\text{T}$ is $r_c \approx 307\,a_0$. In this regime, resonant dipolar couplings mediated by $\sigma_{\pm}\sigma_{\pm}$ and $\sigma_{\mp}\sigma_{\pm}$ transitions dominate, leading to a strong dipolar interaction that scales approximately linearly with $M$
$$
C_3(M) \approx 0.55 M + 0.37 \sqrt{M(M+1)} + 0.28\quad\text{GHz}\cdot\mu\text{m}^3,
$$
enabling tunable long-range interactions. In contrast, off-resonant channels mediated by $\sigma_{-}$ or $\pi$ transitions are highly suppressed due to large energy defects. Notably, the interaction strength can be further enhanced by applying an external electric field, which polarizes the electron cloud relative to the ionic core, generating a strong permanent dipole moment. This configuration offers a promising platform for quantum technological applications that exploit strongly interacting, ultra-long-lived circular rLandau states.

\section{Magnetic Cage}

 One of the key advantages of utilizing Rydberg-Landau states is their inherent ability to suppress the ionization rate, which is achieved by the strong magnetic fields. In particular, the quadratic magnetic potential provides radial confinement, acting effectively as a {\it magnetic cage} or analogous to concepts commonly used in plasma physics literature, {\it magnetic bottle} \cite{Col66,Muc12}. This radial confinement enables the application of higher laser intensities, crucial for coherent population rotations in multi-photon excitations, and substantially improves the interaction-to-loss ratio in Rydberg dressing schemes \cite{Kha24,Kha18,Kha16,Khaz21,Shi24,Khaz21,Shi24}.

\begin{figure}
\centering
    \includegraphics[trim=0 180 510 175, clip, width=\linewidth]{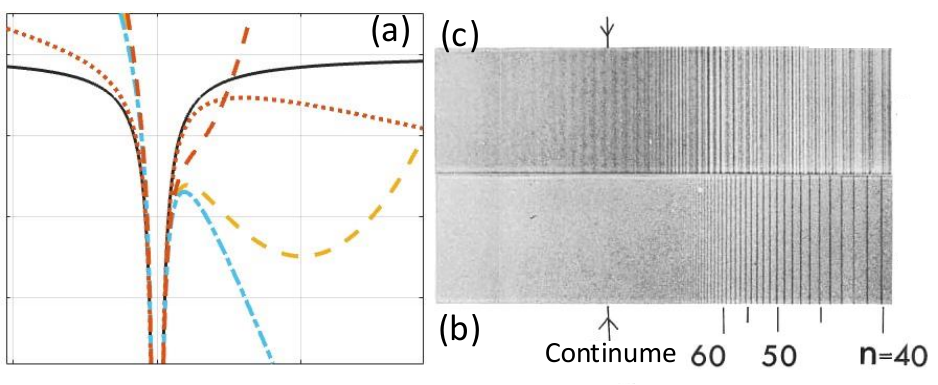}
\caption{{\bf Magnetic Cage for suppressing the ionization. (a)} Considering the coulomb interaction (solid black line), the stark shift caused by the electric field in an electromagnetic wave could cause tunneling ionization for $\gamma<1$ (dotted red line) or over the barrier ionization for $\gamma<<1$ (dashed-dotted line), where $\gamma$ is the Keldish parameter. Adding a magnetic field, the quadratic potential acts as a magnetic cage that prevents ionization (dashed lines - orange for weak and red for strong magnetic field). {\bf  (b,c)} The quasi-Landau states have been observed in the spectroscopy of highly excited Ba atoms under a 5.5T magnetic field. (© AAS. Reproduced with permission \cite{Gar69}). {\bf (b)} zero-field Rydberg spectrum, {\bf (c)} The quasi-Landau states extend into the ionization continuum of {\bf (b)}.
 }\label{Fig_ionization}
\end{figure}

In the absence of a magnetic field, an electron under laser excitation experiences both the Coulomb potential and a Stark shift from the laser field, which lowers the Coulomb barrier, thus facilitating electron tunneling or direct ionization. However, when a strong magnetic field is applied along the laser propagation direction, the electron's dynamics fundamentally change. Classically, the electron is subjected to a harmonic-like magnetic confinement potential in transverse directions due to the Lorentz force, see Fig.~\ref{Fig_ionization}a. Quantum mechanically, this phenomenon manifests as discrete Landau quantization, replacing the previously continuous energy spectrum, see Fig.~\ref{Fig_ionization}b.

Without a magnetic field, ionization involves transitions into a continuous band of free-electron states. Introducing a strong magnetic field alters this scenario, quantizing transverse electron motion into discrete Landau levels spaced by \(\hbar \omega_c\). Consequently, electron escape via ionization now requires discrete transitions into these higher Landau levels, substantially increasing the ionization barrier and significantly suppressing the ionization probability.

Mathematically, this ionization suppression can be explained using Fermi's Golden Rule. The ionization rate \(W\) from an atomic state \(\lvert i \rangle\) to a set of final states \(\lvert f \rangle\) under a circularly polarized light is given by:
$$
W = \sum_{f} \bigl|\langle f | \hat{\sigma}_{\pm} | i\rangle\bigr|^2 \delta(E_f - E_i - \hbar \omega).
$$
Without a magnetic field, the sum over final states becomes an integral over continuous three-dimensional momentum space, enhancing transition probabilities due to a large density of accessible states. In contrast, under strong magnetic fields, the final-state manifold significantly reduces as the continuous transverse momentum space is replaced by discrete Landau levels, with fewer states accessible per energy interval. Thus, the effective density of states in Fermi’s Golden Rule decreases, leading directly to lower ionization rates.


In summary, the discrete quantization of transverse motion into Landau levels induced by a strong magnetic field effectively reduces the density of final states and suppresses the overlap integral, greatly diminishing ionization rates compared to the continuous-state scenario without a magnetic field. This mechanism elucidates why employing Rydberg-Landau states under strong magnetic fields provides a powerful method for ionization suppression, thereby enhancing the effectiveness of laser-induced quantum manipulation schemes.

\section{Experimental Realization}

The level scheme used for exciting rLandau states in \(^{87}\text{Rb}\) is illustrated in Fig.~\ref{Fig_LevelScheme}. We employ a two-photon excitation scheme through the intermediate \(6P_{3/2}\) state, driven by lasers at wavelengths of 420 nm and 1010 nm, characterized by Rabi frequencies \(\Omega_1\) and \(\Omega_2\), respectively. 

When the intermediate state detuning \(\Delta = \omega_1 - \omega_{ps}\) is large compared to the hyperfine structure and natural linewidth of the intermediate \(\ket{p}\) state, the effective two-photon Rabi frequency can be approximated by \(\Omega_{\text{eff}}=\Omega_1\Omega_2/(2\Delta)\). The single-photon Rabi frequencies are explicitly given by:
\[
\Omega_{1,2} = \frac{|e| E_{1,2}}{\hbar}\,\langle p|\mathbf{r}\cdot\boldsymbol{\epsilon}_{1,2}|s\rangle,
\]
where \(E_{1,2}\) and \(\boldsymbol{\epsilon}_{1,2}\) represent the amplitude and polarization vectors of the applied fields, respectively. The radial transition dipole integrals for \(^{87}\text{Rb}\) are known for the \(5S \rightarrow 6P\) transition, with values of \(\langle r\rangle_{5S}^{6P_{3/2}}=0.528\,a_0\) and \(\langle r\rangle_{5S}^{6P_{1/2}}=0.235\,a_0\). For transitions from the \(6P\) and \(5P\) intermediate states to the rLandau state \(\ket{N=100,N_\ell=0,M=0}\), we numerically obtain the dipole integrals as \(\langle r\rangle_{6P}^{N=100,N_\ell=0,M=0}=3.5\times10^{-4}\,a_0\) and \(\langle r\rangle_{5P}^{N=100,N_\ell=0,M=0}=1.4\times10^{-4}\,a_0\), respectively.

To estimate the achievable Rabi frequency for the \(6P\) to rLandau transition, we consider currently available strong continuous-wave laser systems with output powers of around 50 W \cite{Gou19,Gou19N}, potentially scalable using fiber amplifiers \cite{FiberAmp}. Assuming a Gaussian beam profile with waist radius \(w=1\,\mu\text{m}\), the electric field amplitude at the beam center is calculated from the beam power \(P\) using:
\[
P = \frac{E_0^2}{2\eta}\int \exp\left(-\frac{2r^2}{w^2}\right)2\pi r\,dr,
\]
where \(\eta=377\) is the free-space impedance. For a beam power of 100 W, the achievable Rabi frequency is approximately \(\Omega_2 = eE_0\langle r\rangle_{6P}^{N_\ell=0,M=0}/\hbar\approx 2\pi\times1\,\text{GHz}\). Given the substantially larger dipole moment for the \(5S \rightarrow 6P_{3/2}\) transition, achieving even higher single-photon Rabi frequencies for \(\Omega_1\) is feasible.

A major practical consideration is the partial excitation of the short-lived intermediate state, which introduces spontaneous emission and decoherence. The intermediate states \(6P_{3/2}\) and \(5P_{3/2}\) have short lifetimes (\(\tau_{6P}=129\,\text{ns}\), \(\tau_{5P}=28\,\text{ns}\)). To mitigate spontaneous emission from these states, we use a large intermediate-state detuning \(\Delta\). During a two-photon \(\pi\)-pulse of duration \(t=\pi/|\Omega_{\text{eff}}|\), the probability of spontaneous emission-induced decoherence from the intermediate state is given by \cite{Saf10}:
\[
P_{\text{se}}=\frac{\pi}{4|\Delta|\tau_p}\left(q+\frac{1}{q}\right),\quad\text{where}\quad q=\left|\frac{\Omega_1}{\Omega_2}\right|.
\]

Importantly, the presence of a strong magnetic field (2.5 T) substantially suppresses the ionization rate (see discussion in previous sections), enabling the use of stronger laser intensities. This allows for larger intermediate-state detunings, further reducing the probability of spontaneous emission and significantly enhancing coherence during the excitation process.

\section{Ryd-Landau Excitation revisited}

To elucidate the dynamics underlying rLandau state excitation, we briefly revisit wavepacket dynamics during hydrogenic Rydberg excitation from the \(\ket{5S}\) ground state in rubidium \cite{Alb91}. On picosecond timescales immediately following excitation, the size of the Rydberg wavepacket remains close to that of the original \(\ket{5S}\) state. Subsequently, this wavepacket rapidly expands outward and gradually diffuses into higher-lying Rydberg orbitals on the nanosecond timescale.

In the case of rLandau state excitation, the electron initially experiences the predominantly Coulombic potential near the ionic core. Immediately after the two-photon dipole excitation, the electron occupies standard hydrogenic Coulomb eigenstates \(\ket{nS}\), with the principal quantum number \(n\) determined by the laser frequency. Subsequently, this \(nS\) wavepacket moves outward, guided by the new potential landscape set by the highly excited state. For large principal numbers, the electron enters regions far from the core, where the quadratic magnetic confinement dominates over the Coulomb potential. As a result, the wavepacket redistributes and evolves into the corresponding Rydberg–Landau eigenstates (rLandau states).


{\bf Author contribution} The project was conceived, developed, supervised, and written by M. Khazali. A. Momtaheni contributed in the coding of the rLandau wavefunction and assisted with parts of the selection rule derivations and lifetime calculations in the context of his master's thesis, conducted under the supervision of M. Khazali.

 \end{document}